\documentclass[fleqn,usenatbib]{mnras}

\usepackage{newtxtext,newtxmath}

\usepackage[T1]{fontenc}
\usepackage{ae,aecompl}

\usepackage{graphicx}	
\usepackage{subfigure}

\usepackage{natbib}
\usepackage{subfigure}
\usepackage{color}
\usepackage{verbatim}
\usepackage{multirow}
\usepackage{lscape}
\usepackage{afterpage}

\newcommand{\Teff}{{T_\mathrm{eff}}}
\newcommand{\logg}{{\log g}}
\newcommand{\FeH}{\mathrm{[Fe/H]}}
\newcommand{\XL}{\mathrm{X_{low}}}
\newcommand{\XH}{\mathrm{X_{high}}}
\newcommand{\XLXH}{\mathrm{[X_{low}/X_{high}]}}
\newcommand{\AlFe}{\mathrm{[Al/Fe]}}
\newcommand{\TiFe}{\mathrm{[Ti/Fe]}}
\newcommand{\CoFe}{\mathrm{[Co/Fe]}}
\newcommand{\KSi}{\mathrm{[K/Si]}}

\title[The Metallicity Effect on LDR]{The metallicity effect on line-depth ratios in APOGEE $H$-band spectra}

\author[M. Jian et al.]{
	Mingjie Jian,$^{1}$\thanks{E-mail: mingjie@astron.s.u-tokyo.ac.jp}
	Noriyuki Matsunaga,$^{1}$
	Kei Fukue$^{2}$
	\\
	$^{1}$Department of Astronomy, The University of Tokyo, 7-3-1 Hongo, Bunkyo-ku, Tokyo 113-0033, Japan\\
	$^{2}$Laboratory of Infrared High-resolution spectroscopy (LiH), Koyama Astronomical Observatory, Kyoto-Sangyo University, \\Motoyama,
	Kamigamo, Kita-ku, Kyoto 606-8555, Japan
}

\date{Accepted XXX. Received YYY; in original form ZZZ}

\pubyear{2018}

\begin{document}
\label{firstpage}
\pagerange{\pageref{firstpage}--\pageref{lastpage}}
\maketitle
	
	
\begin{abstract}
		Ratios of carefully selected line depths are sensitive to stellar effective temperature ($\Teff$).
		Relations established between line-depth ratio (LDR) and $\Teff$ allow one to determine $\Teff$ precisely.
		However, LDRs can also depend on metallicity and abundance ratios, which can limit the accuracy of the LDR method unless such effects are properly taken into account.
		We investigate the metallicity effect using $H$-band spectra and stellar parameters published by the APOGEE project. 
		We clearly detected the effects of metallicity and abundance ratios; $\Teff$ derived from a given LDR depends on the metallicity, 100--800\,K\,dex$^{-1}$, and the dependency on the abundance ratios, 150--1000\,K\,dex$^{-1}$, also exists when the LDR involves absorption lines of different elements. 
		For the 11 line pairs in the $H$-band we investigated, the LDR--$\Teff$ relations with abundance-related terms added have scatters as small as 30--90\,K within the range of $3700 < \Teff < 5000$\,K and $-0.7 < \FeH < +0.4$\,dex.
		By comparing the observed spectra with synthetic ones, we found that saturation of the absorption lines at least partly explains the metallicity effect.
\end{abstract}
	
	
	
	
	\section{Introduction}
	
	Determining the effective temperature ($\Teff$) of a star is a fundamental step in stellar spectral analysis.
	There are various methods to determine $\Teff$, such as using colour--$\Teff$ relations (e.g. \citealt{Bessell_1998}) and combining interferometric diameters and bolometric fluxes (e.g. \citealt{Heiter_2015}).
	Many of these methods, however, work only for stars whose foreground extinction are not severe.
	Unlike the methods using colours or interferometric data, absorption lines in continuum-normalised spectra are not affected by interstellar reddening or extinction except rare and weak diffuse interstellar bands (\citealt{IR_band_1}; \citealt{IR_band_2}).
	Thus, a line-depth ratio (LDR), which measures a ratio of two metallic lines with different excitation potentials (EPs), can be a useful indicator of $\Teff$ for a wide range of the Milky Way disc in which the interstellar extinction tends to be a problem (\citealt{MN2017}).
	However, most previous studies on the LDR method considered only optical spectra.
	For example, such works applied to giants, which is our targets,  include: \citet{strassmeier_temperature_2000}, \citet{gray_line-depth_2001}, and \citet{kovtyukh_determinations_2006}.
	Recently, \citet{fukue_line-depth_2015} found nine LDR--$\Teff$ relations using $H$-band spectra of eight stars (mainly giants), and \citet{taniguchi} found 81 relations using $YJ$-band spectra of nine giants.
	Their relations are the first found in these wavelengths, but the limited numbers of their samples left many points to be addressed, e.g. more precise calibration and metallicity effects on the LDRs.
	
	It has been expected, to the first approximation, that the LDRs only weakly depend on metallicity and other parameters.
	\citet{sasselov_accurate_1990} investigated LDRs combining C I  and Si I lines around 1.1\,$\mu \mathrm{m}$ and found that the dependency on gravity and microturbulence is small (see their fig.~1).
	However, by studying five LDRs in optical spectra of 92 red giants with a spread in metallicity, \citet{gray_line-depth_2001} found that there is a significant scatter around the relations between LDR and colour (a proxy for temperature), and that the deviation is correlated with metallicity as well as absolute magnitude (a proxy for surface gravity).
	Moreover, although the numbers of stars studied in \citet{fukue_line-depth_2015} and \citet{taniguchi} are limited, their studies show offsets of metal-poor stars in some LDR--$\Teff$ relations.	
	It is important to characterize the metallicity effect, and such an effect, if any, would have to be taken into account in the determination of $\Teff$.
	
	In this work, we investigate the effects of metallicity and abundance ratios on LDR--$\Teff$ relations using a large number of $H$-band spectra collected in the APOGEE survey \citep{APOGEE}.
	The APOGEE spectra cover eight pairs of absorption lines among the nine reported by \citet{fukue_line-depth_2015}.
    The spectra also include wavelengths in the gaps between echelle orders of the spectra used by \citet{fukue_line-depth_2015}, which allows us to search for new line pairs useful for the LDR method.
	We describe the selection and analysis of the APOGEE data in Section~\ref{sec:data}.
	In Section~\ref{sec:result}, we first derive LDR--$\Teff$ relations for solar-metal objects and then extend the metalliticy range to derive the relations including metallicity terms.
	The cause of the metallicity effect is discussed in Section~\ref{sec:discussion}, and the summary is given in Section~\ref{sec:summary}.
	
	\section{Data and Analysis}
	\label{sec:data}
	
	\subsection{APOGEE Catalogue and $H$-band Spectra}
	
	The main targets of the APOGEE survey are red giants selected for the purpose of studying stellar populations spread over various regions of the Milky Way (i.e.~bulge, disk, and halo).
	Their Data Release 14 (DR14) provides us with temperatures and other parameters of individual objects obtained by APOGEE Stellar Parameters and Chemical Abundances Pipeline (ASPCAP, \citealt{garcia_perez_aspcap:_2016}) together with individual $H$-band spectra ($R \, {\sim} 22,500$).
	The spectra cover a wavelength range from $1.514$ to $1.696$\,$\mu$m with two gaps at $1.581$--$1.585$\,$\mu$m and $1.644$--$1.647$\,$\mu$m.
	Using these homogeneous $H$-band spectra of numerous objects with the parameters estimated, the temperature scales in \citet{fukue_line-depth_2015} can be tested with a significantly larger sample.
	Moreover, it is possible to investigate the metallicity effect on the LDR--$\Teff$ relations by comparing the LDRs at different metallicities.
	
	\subsection{Selection of APOGEE sample}
	\label{sec:selection}
	
	For our analysis, 17,459 spectra were selected among 277,371 included in DR14 on the following criteria:
	
	\begin{enumerate}
		\item Signal-to-noise ratio (${\tt S/N}$) given in their catalogue is higher than $300$.
		\item The data quality flags, {\tt STARFLAG}, and two flags inside {\tt ASPCAPFLAGS} (i.e. {\tt STAR\_WARN} and {\tt STAR\_BAD}) are 0 (which means no warning was given). These flags, if non-zero values are given, indicate at least one of the various shortcomings such as low S/N, poor matches to synthetic spectra, being too close to the limit of available grid points for models, and so on\footnote{https://www.sdss.org/dr14/algorithms/bitmasks/}. 
		\item Six abundances relevant to our study, $\FeH$, $\AlFe$, $\CoFe$, $\mathrm{[K/Fe]}$, $\mathrm{[Si/Fe]}$ and $\TiFe$, are available together with errors, and the corresponding flags inside {\tt ELEMFLAG}\footnote{https://www.sdss.org/dr14/irspec/abundances/} are 0 (no issue in the abundance determination) or 512 ({\tt OTHER\_WARN}, other warning condition). Most of the stars with the flag 512 have low or high temperatures, $\Teff \precsim 4100$\,K or $\Teff \succsim 5100$\,K, but they show the same trend in the abundance--metallicity diagrams as the stars with the flag 0. Although abundances for these low- or high-$\Teff$ stars may not have been properly calibrated, including them allows us to enlarge the temperature range for this work. 
	\end{enumerate}

\begin{figure}
		\centering
		\includegraphics[width=\columnwidth]{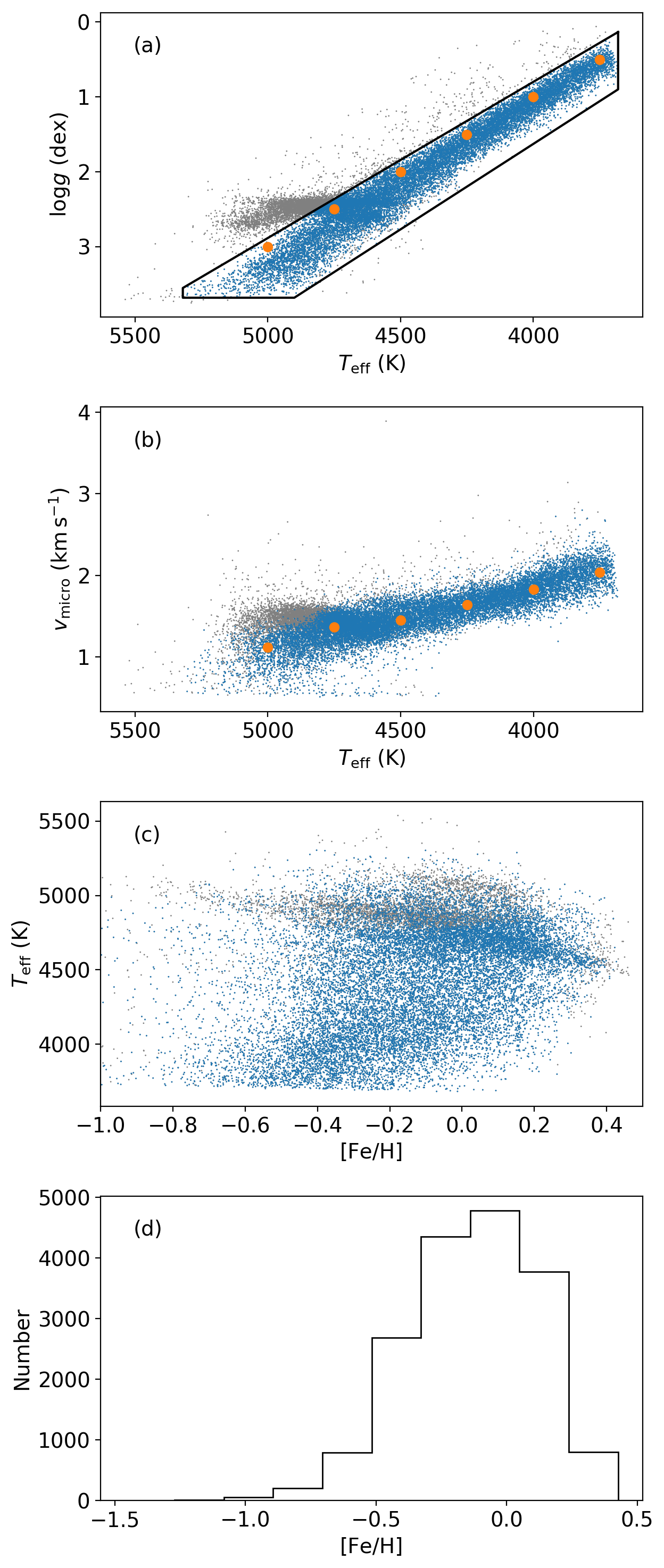}
		\caption{Parameters of the 17,459 APOGEE targets we selected in Section~\ref{sec:selection}. (a) Hertzsprung-Russell diagram, (b) microturbulence plotted against $\Teff$, (c) $\Teff$ plotted against [Fe/H], and (d) histogram of [Fe/H]. The blue points indicate the stars we selected based on the black box shown in the panel (a). The orange points in (a) and (b) indicate the $\logg$ and microturbulence values used for generating synthetic spectra (see Section~\ref{sec:discussion}).}
		\label{fig:FeH}
	\end{figure}

\begin{figure}
		\centering
		\includegraphics[width=\columnwidth]{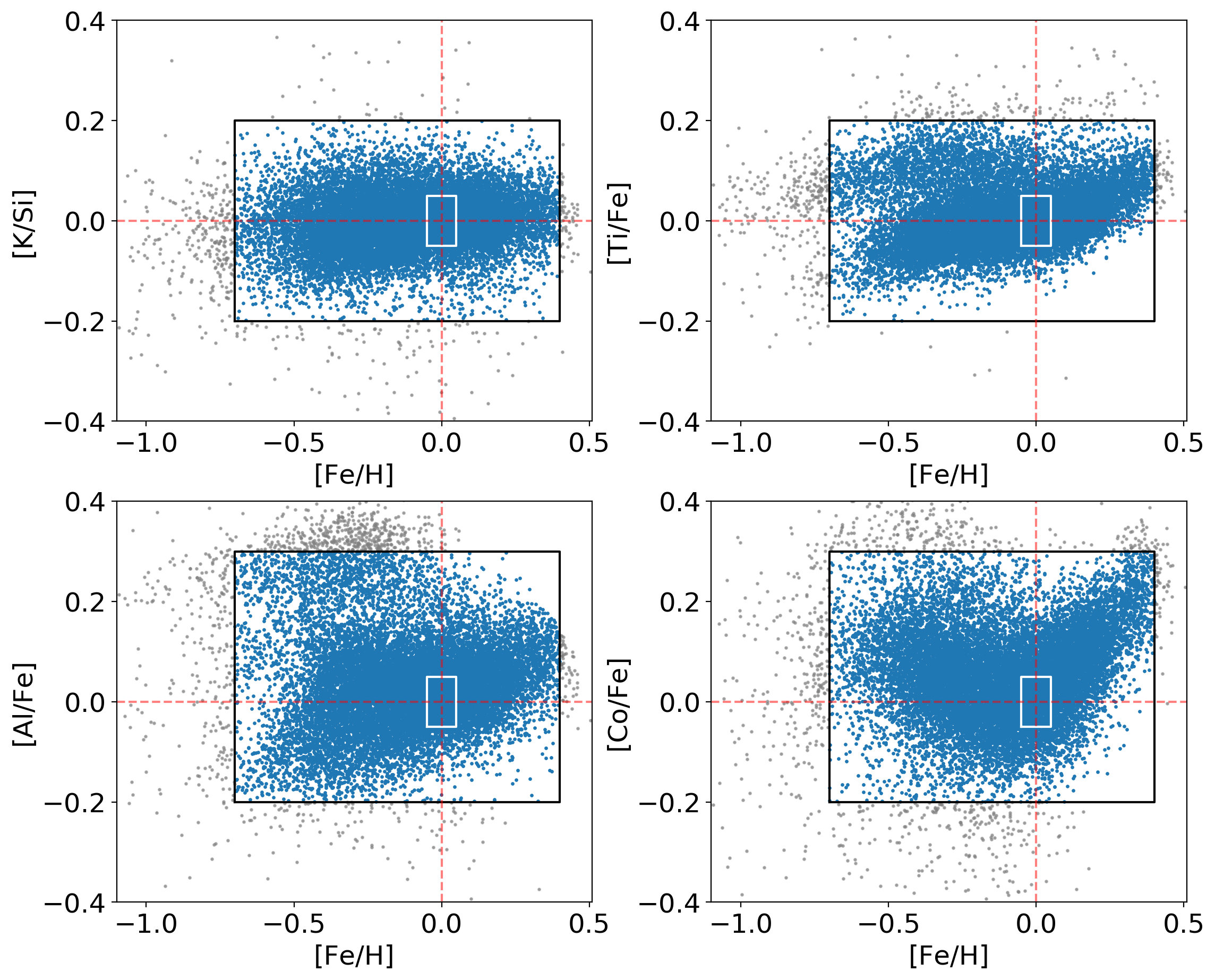}
		\caption{Abundance ratios, [K/Si], [Ti/Fe], [Al/Fe], and [Co/Fe], are plotted against the metallicity for the 17,459 targets selected in Section~\ref{sec:selection}. Sources rejected in Fig.~\ref{fig:FeH} are not included in this plot, and those indicated by grey points in this plot are not used in the analysis either. Those inside the white and black boxes are used for the fits in Section~\ref{sec:solar-fit} and~\ref{sec:all-metal-fit}, respectively.}
		\label{fig:XFe}
	\end{figure}

    Fig.~\ref{fig:FeH} plots the stellar parameters of the APOGEE sample, $\Teff$, $\logg$, microturbulence, and $\FeH$.
	The metallicity is well spread between $-0.7$ and $+0.4$\,dex for the stars with a wide range of temperature, $3700 < \Teff < 5000$\,K (roughly corresponding to the range of spectral types G6--M2). 
	It should be noted that red giants show a correlation between $\Teff$ and the gravity, $\logg$, and so does the APOGEE sample as illustrated in Fig.~\ref{fig:FeH} (a) (see also fig.~20 in \citealt{APOGEE}).
	It is therefore difficult to investigate the gravity effect, if any, independently.
	As a result, the LDR--temperature scale we study here should be applied to red giants which fall more-or-less on the branch formed by the APOGEE sample.
    We limited the samples within the black box in Fig.~\ref{fig:FeH} (a) to remove red clump stars as well as some outliers which are clearly deviated from the red giant branch.
	Fig.~\ref{fig:XFe} plots relevant abundance ratios against the metallicity of our sample (after the red clump stars and outliers removed).
	The wide ranges of the metallicity and the abundance ratios allow us to study the metallicity effects on LDR--$\Teff$ relations.
	
	\subsection{Measurement of the LDRs}
	
	We adopt the line pairs used in \citet{fukue_line-depth_2015} and also search for new line pairs in the APOGEE spectra.
	Among the nine line pairs in \citet{fukue_line-depth_2015}, eight (IDs 2 -- 9) are within the APOGEE wavelength range, while the K I line of the pair (1) at $15163.09$\,\AA~is outside.
	However, we reject the line pair (7) for the following reason.
	The line list of \citet{melendez_oscillator_1999} has two Fe I lines with different EPs, $6.38$ and $2.18$\,eV, at the same wavelength, $\lambda=16225.64$\,\AA.
	These lines are also very close to each other ($\Delta \lambda < 0.02$\,\AA) in the list adopted by APOGEE (\citealt{shetrone_sdss-iii_2015}).
	Our simulation making use of synthetic spectra confirmed that the two lines have significant contributions with different sensitivity to temperature: the high EP line is dominant at high $\Teff$ (${\sim}6000$\,K), while the two become comparable at lower $\Teff$.
	Such a situation leads to a complicated response to $\Teff$ and possibly to other parameters.
	In fact, the line pair (7) gives the largest scatter around the fitted LDR--$\Teff$ relation in \citet{fukue_line-depth_2015}, and we also confirmed this with the APOGEE dataset.
	Thus we exclude this line pair in the following analysis.
    
	On the other hand, based on the list of \citet{shetrone_sdss-iii_2015}, we searched for new pairs of lines with significantly different EPs which give tight LDR--$\Teff$ relations.
	All atomic lines with EP $< 5$\,eV were classified as low EP lines, but we excluded those which have other atomic lines within ~2\,\AA. 
	We then tried to find high EP lines (EP $> 5$\,eV) which have no other atomic lines within 2\,\AA~but are close to each of the low EP lines. 
	Four new pairs, IDs 10--13, were identified (Table~\ref{tab:linelistn}). 
	
	The spectra in the APOGEE DR14 are in vacuum and rest wavelength, i.e.~shifts caused by stellar radial velocities have been corrected.
	We made the transformation to air wavelength (using the formula in \citealt{vacuum_air}) which is directly comparable with wavelengths in \citet{fukue_line-depth_2015}.

	\begin{table*}
		\centering
		\caption{Line pairs used in our LDR--$\Teff$ relations. Parameters of each line are taken from \citet{shetrone_sdss-iii_2015} with $\lambda$ transformed into the air wavelength scale.}
		\begin{tabular}{ c c c c c c c c c c}
			\hline
			& \multicolumn{4}{c}{Low-excitation Line} & & \multicolumn{4}{c}{High-excitation Line} \\
			\cline{2-5}\cline{7-10}\\
			ID & $\XL$ & $\lambda$ (\AA) & EP(eV) & log $gf$ (dex) & & $\XH$ & $\lambda$ (\AA) & EP(eV) & log $gf$ (dex)\\\hline
			(2) & K  I & 15168.38 & 2.67 &  0.52 && Si I & 15376.83 & 6.22 & $-$0.72\\
			(3) & Fe I & 15194.49 & 2.22 & $-$4.73 && Fe I & 15207.53 & 5.39 & 0.12\\
			(4) & Ti I & 15543.76 & 1.88 & $-$1.16 && Fe I & 15591.49 & 6.36 &  0.97\\
			(5) & Ti I & 15602.84 & 2.27 & $-$1.63 && Fe I & 16040.66 & 5.87 & 0.07\\
			(6) & Ti I & 15715.57 & 1.30 & $-$1.59 && Fe I & 15621.65 & 5.54 &  0.26\\
			(8) & Co I & 16757.60 & 3.41 & $-$1.85 && Fe I & 16316.32 & 6.28 &  0.92\\
			(9) & Al I & 16763.36 & 4.09 & $-$0.51 && Fe I & 16517.23 & 6.29 & 0.52\\
			(10) & Fe I & 15490.34 & 2.20 & $-$4.79 && Fe I & 15534.24 & 5.64 & $-$0.34\\
			(11) & Ti I & 15698.98 & 1.89 & $-$2.11 && Fe I & 15723.59 & 5.62 & $-$0.01\\
			(12) & Ti I & 16401.51 & 2.33 & $-$2.11 && Fe I & 16394.39 & 5.96 & 0.00\\
			(13) & Fe I & 15611.15 & 3.41 & $-$3.18 && Fe I & 15604.22 & 6.24 & 0.65\\
			\hline
		\end{tabular}
		\label{tab:linelistn}
	\end{table*}
	
	Although the pseudo-continuum normalised spectra are given in DR14, the continuum level tends to be significantly higher than the unity, and further normalisation is required for our purpose.
	For each part of the spectra separated by the gaps, we ran the $\textsc{IRAF~continuum}$ task 10 times using the cubic spline function with iterative rejections of the pixels more than $1.6 \sigma$ below the fit where $\sigma$ is the residual at each iteration. 
	We created histograms of count distribution of the re-normalised spectra, and the full widths at half maximum for the continuum part were converted into errors of continuum determination ($\sigma_\mathrm{con}$) assuming that the errors follow Gaussian.

	Then four or five points around a given line centre were fitted with Gaussian and parabola curves.
    The Gaussian-Hermite profile, identified as the line spread function of APOGEE spectra by \citet{garcia_perez_aspcap:_2016}, was also applied to fit nine points around the centre.
    These give us three estimates of the depth.
	If a depth was smaller than $3\sigma_\mathrm{con}$, the measurement was discarded.
	We also rejected the measurement if the distance between the line's rest wavelength and the minimum of the fitted curve was larger than a pixel (${\sim}0.2$\,\AA) or if the measured depth had a statistical error larger than $0.05$.
    Such errors can be seen in lines for which blends with other lines are too strong or bad pixels skew the line profile.
	Approximately $20\%$ of the measurements were rejected.
	The line depths measured from the three profiles are similar to each other; the systematic differences between the depth measurement with the different profiles are approximately 1\% for all the line pairs, and the standard deviations of the differences are typically of the order of 2\%.
    Considering the conclusion by \citet{strassmeier_temperature_2000} and that the Gaussian-Hermite profile requires more points which may be more affected by blends, we use the depths measured by the parabola fitting in the following analysis unless otherwise mentioned.
    The impacts of blending lines around the selected lines will be discussed in Section~\ref{sec:line_blend}. 
	
	\section{Results}
	\label{sec:result}

	\subsection{LDR--$\Teff$ relations of the solar-metal sample}
	\label{sec:solar-fit}
	
\begin{figure*}
		\includegraphics[width=2\columnwidth]{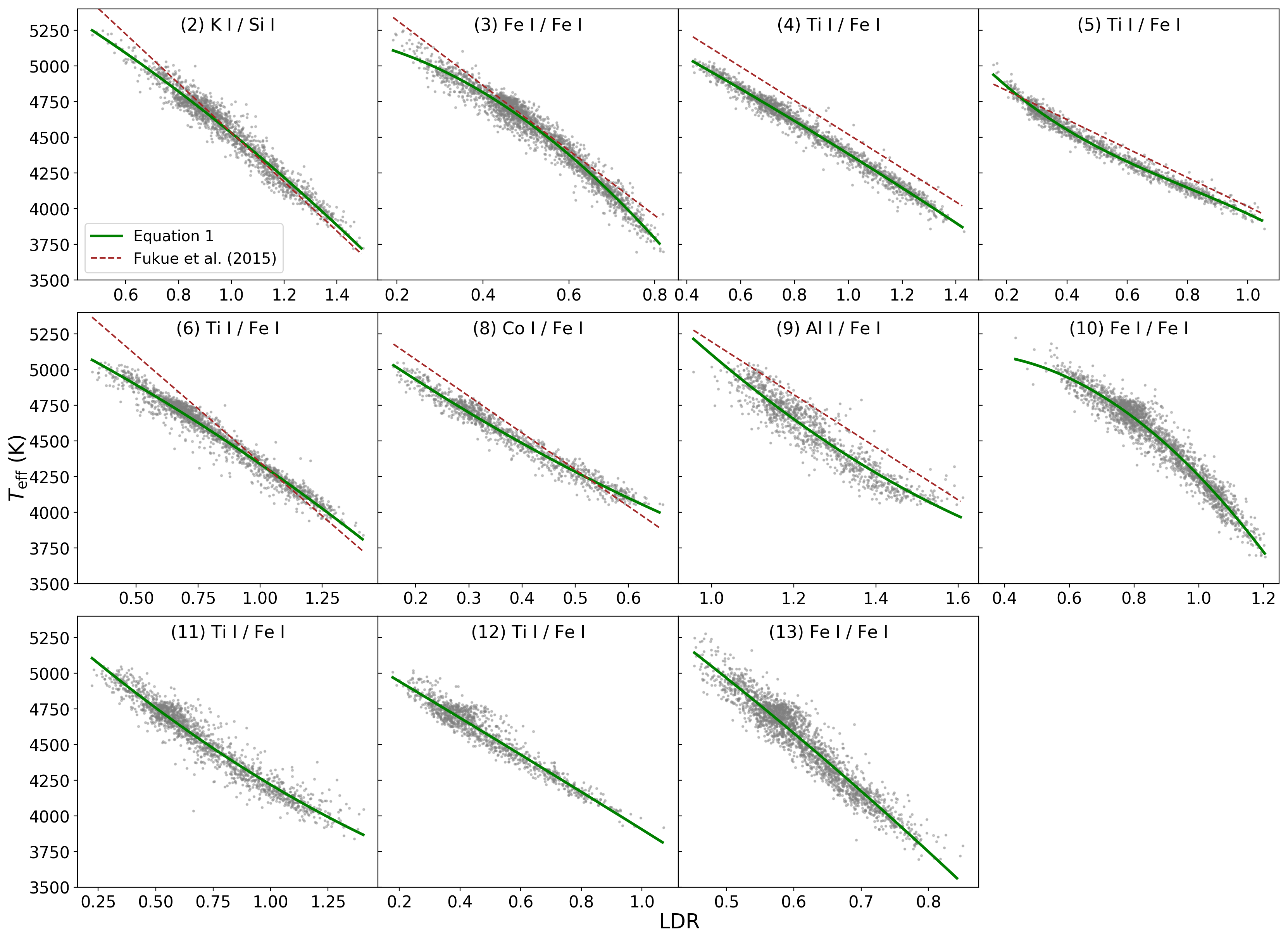}
		\caption{LDR--$\Teff$ relations of the solar-metal stars. The fitted relations we obtained (Table~\ref{tab:solarfit}) are indicated by green curves, while brown dashed lines indicate the relations in \citet{fukue_line-depth_2015}.}
		\label{fig:solarfit}
	\end{figure*}

	First, we consider only solar-metal stars, i.e. $-0.05 < \FeH < 0.05$~dex.
	The abundance ratio of the element for the low EP line to that for the high EP line, $\XLXH$, was also required to be around the solar, zero, within $\pm$0.05 dex.
	Fig.~\ref{fig:solarfit} plots the obtained LDRs versus $\Teff$. It is clear that linear fits are not sufficient for some relations, and we obtained least-squares fits in the form of	
	\begin{equation}
	\Teff = a (r-r_0) ^2 + b(r-r_0) + c,
	\label{eq:solar}
	\end{equation}
	where $r$ is the LDR (the ratio of low to high EP line depths, $d_{\mathrm{low}} / d_{\mathrm{high}}$) and $r_0$ is the intermediate LDR value for each relation.
	The obtained coefficients are given together with $r_0$ in Table~\ref{tab:solarfit}.
	The third-order polynomial formula is also tested with the same dataset but the improvements are not significant except for the line pair (5), and thus we adopt the third-order term, $a'(r-r_0)^3$, only for the pair (5). 
	Fig.~\ref{fig:solarfit} draws the obtained relations as well as the linear relations in \citet{fukue_line-depth_2015} where available.
	The new relations we found, i.e.~(10)--(13), are reasonably tight, suggesting that they are as useful as the previously reported line pairs.
	The mean residual value of our relations is ${\sim}60$\,K,  smaller than the residuals found by \citet{fukue_line-depth_2015}, ${\sim}150$\,K.
	Their relations tend to be slightly offset from the distribution of our measurements.
	The moderate offsets, $100$--$200$\,K, are still at the same order of the residuals around the fitted relations which \citet{fukue_line-depth_2015} obtained for the eight calibrating stars; their result may also be affected by the metallicity as we discuss in the next section.
    
	\begin{table*}
		\centering
		\caption{LDR--$\Teff$ relations in the form of equation~\ref{eq:solar} for the solar-abundance samples. For all the line pairs, $\FeH$ is limited within $[-0.05, 0.05]$. The residual around the fits, $\sigma$, and the number of the data points used, $N$, are given for each line pair. For the line pair (5), we considered the third-order term, $a'(r-r_0)^3$, which is included in the third column.}
		\begin{tabular}{ c c c c c c c c }
			\hline
			ID & $r_0$ & $a$ & $b$ & $c$ & $\sigma$(K) & Abundance range & $N$ \\\hline
            (2) & $1.0$ &   $-271 \pm     38$ &  $-1507 \pm     9$ &   $4531 \pm 2.1$ &   $64$ & $-0.05< \mathrm{[K/Si]} < 0.05$  &  $1626$\\
            (3) & $0.5$ &  $-1894 \pm     67$ &  $-2179 \pm      11$ &   $4616 \pm 1.6$ &   $60$ & -- & $2395$\\
            (4) & $0.8$ &  $-89 \pm     20$ &  $-1139 \pm      6$ &   $4616 \pm 1.3$ &   $41$ & $-0.05< \mathrm{[Ti/Fe]} < 0.05$ & $1715$\\
            (5) & $0.5$ &  $ (-835 \pm     103)(r-r_0) + 756 \pm     38$ &  $-1110 \pm      8$ &   $4434 \pm 1.6$ &   $34$ & $-0.05 < \mathrm{[Ti/Fe]} < 0.05$ & $1203$\\
            (6) & $0.8$ &  $-185 \pm     23$ &  $-1128 \pm      6$ &   $4572 \pm 1.7$ &   $53$ & $-0.05< \mathrm{[Ti/Fe]} < 0.05$ & $1729$\\
            (8) & $0.4$ &  $764 \pm     111$ &  $-2075 \pm      13$ &   $4484 \pm 2.1$ &   $47$ & $-0.05< \mathrm{[Co/Fe]} < 0.05$ & $1042$\\
            (9) & $1.2$ &  $940 \pm     163$ &  $-2075 \pm      36$ &   $4654 \pm 3.4$ &   $94$ & $-0.05< \mathrm{[Al/Fe]} < 0.05$ & $1106$\\
            (10) & $0.8$ &  $-1602 \pm     65$ &  $-1709 \pm      14$ &   $4662 \pm 2.0$ &   $72$ & -- & $2170$\\
            (11) & $1.0$ &  $229 \pm     29$ &  $-964 \pm      15$ &   $4219 \pm 2.8$ &   $73$ & $-0.05< \mathrm{[Ti/Fe]} < 0.05$ & $1642$\\
            (12) & $0.5$ &  $-34 \pm     62$ &  $-1290 \pm      13$ &   $4559 \pm 2.5$ &   $61$ & $-0.05< \mathrm{[Ti/Fe]} < 0.05$ & $1107$\\
            (13) & $0.6$ &  $-956 \pm     293$ &  $-3962 \pm      30$ &   $4580 \pm 2.3$ &   $87$ & -- & $2312$\\
			\hline
		\end{tabular}
		\label{tab:solarfit}
	\end{table*}

	\subsection{LDR--$\Teff$ relations with metallicity terms}
	\label{sec:all-metal-fit}
	
\begin{figure*}
		\includegraphics[width=2\columnwidth]{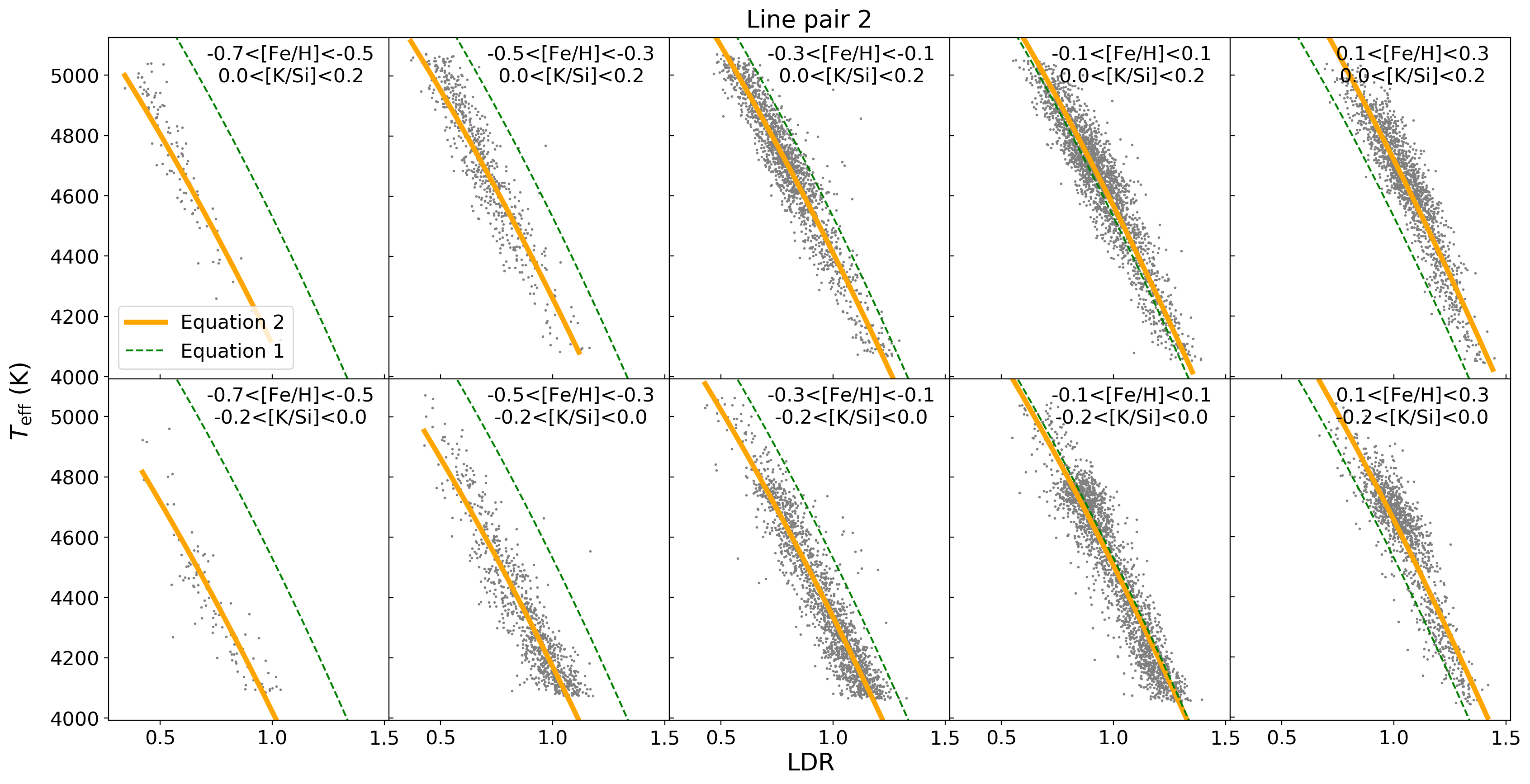}
		\caption{LDR--$\Teff$ relations for samples within different abundance ranges. The gray dots indicate objects within the labelled abundance range and the orange curve shows the cut of equation~\ref{eq:wide} at the mean abundance of the objects included in each panel. The green line indicates the relation for the solar-metal sample (equation~\ref{eq:solar}). The plot presented here is for the line pair (2) with K I and Si I lines, but similar plots for all the line pairs are available as Supporting Information.}
		\label{fig:r1}
	\end{figure*}

	In order to investigate the metallicity effect on the LDR--$\Teff$ relations, we here consider least-squares fits in the form of
	\begin{multline}
	\label{eq:wide}
	\Teff = a (r-r_0) ^2 + b(r-r_0) + c + \\ d\FeH + e\FeH(r-r_0) + f\XLXH
	\end{multline}
	to the samples inside the black boxes in Fig.~\ref{fig:XFe}.
	The last term is omitted for the line pairs with two Fe lines.
	Similarly to the previous section, an extra term $a'(r-r_0)^3$ is added for the line pair (5).
	As given in Table~\ref{tab:allfit}, all the relations have significant abundance-dependent terms.
	For example, Fig.~\ref{fig:r1} illustrates how the LDR--$\Teff$ relation depends on $\FeH$ and $\KSi$ in case of the line pair (2).
	The relation for the given abundance, the orange curve, in each panel shows a systematic offset from the relation of the solar-metal sample, the green curve, depending on $\FeH$ and $\KSi$.
	
	\begin{table*}
		\centering
		\caption{Similar to Table~\ref{tab:solarfit} but for LDR--$\Teff$ relations in the form of equation~\ref{eq:wide} for the wide-abundance samples. For all the line pairs, $\FeH$ is limited within $[-0.7, 0.4]$. The same intermediate LDRs, $r_0$, as those given in Table~\ref{tab:solarfit} are adopted.}
		\begin{tabular}{c c c c c c c c c c}
			\hline
			ID  & $a$ & $b$ & $c$ & $d$ & $e$ & $f$ & $\sigma$(K) & Abundance range & N \\\hline
            (2) &    $-235 \pm   20$ &  $-1481 \pm    4$ & $4536 \pm 0.7$ &  $851 \pm 4$ &  $51 \pm  22$ &  $782 \pm  11$ &   $62$ & $-0.2 < \mathrm{[K/Si]} < 0.2$ & 11769\\
            (3) &    $-942 \pm   35$ &  $-2144 \pm    6$ & $4600 \pm 0.7$ &  $319 \pm 3$ &  $-711 \pm  24$ &  $-$ &   $60$ & $-$ & 12889\\
            (4) &    $67 \pm   7$ &  $-1155 \pm    2$ & $4614 \pm 0.5$ &  $160 \pm 2$ &  $-492 \pm  9$ &  $427 \pm  7$ &   $43$ & $-0.2 < \mathrm{[Ti/Fe]} < 0.2$ & 12800\\
            (5) &    $(-822 \pm   56)(r-r_0) + 711 \pm    14$ & $-1111 \pm 4$ &  $4438 \pm 0.5$ &  $304 \pm  2$ &  $-574 \pm  11$ &   $572 \pm  6$ & $31$ & $-0.2 < \mathrm{[Ti/Fe]} < 0.2$ & 8651\\
            (6) &    $-40 \pm   9$ &  $-1124 \pm    3$ & $4574 \pm 0.6$ &  $269 \pm 2$ &  $-295 \pm  11$ &  $268 \pm  8$ &   $54$ & $-0.2 < \mathrm{[Ti/Fe]} < 0.2$ & 12915\\
            (8) &    $1005 \pm   34$ &  $-2078 \pm    4$ & $4477 \pm 0.7$ &  $429 \pm 2$ &  $-295 \pm  23$ &  $508 \pm  5$ &   $48$ & $-0.2 < \mathrm{[Co/Fe]} < 0.3$ & 11937\\
            (9) &    $952 \pm   42$ &  $-2022 \pm    12$ & $4655 \pm 1.1$ &  $112 \pm 6$ &  $14 \pm  34$ &  $1056 \pm  11$ &   $91$ & $-0.2 < \mathrm{[Al/Fe]} < 0.3$ & 11556\\
            (10) &    $-639 \pm   28$ &  $-1730 \pm    7$ & $4640 \pm 0.9$ &  $251 \pm 4$ &  $-1038 \pm  25$ &  $-$ &   $71$ & $-$ & 10936\\
            (11) &    $293 \pm   13$ &  $-892 \pm    8$ & $4240 \pm 1.4$ &  $458 \pm 7$ &  $-28 \pm  18$ &  $162 \pm  13$ &   $78$ & $-0.2 < \mathrm{[Ti/Fe]} < 0.2$ & 11834\\
            (12) &    $34 \pm   30$ &  $-1277 \pm    6$ & $4560 \pm 1.1$ &  $125 \pm 4$ &  $432 \pm  27$ &  $446 \pm  14$ &   $68$ & $-0.2 < \mathrm{[Ti/Fe]} < 0.2$ & 7761\\
            (13) &    $912 \pm   149$ &  $-3868 \pm    15$ & $4566 \pm 1.0$ &  $707 \pm 4$ &  $-2189 \pm  71$ &  $-$ &   $87$ & $-$ & 12124\\\hline
		\end{tabular}
		\label{tab:allfit}
	\end{table*}
	
\begin{figure*}
		\includegraphics[width=2\columnwidth]{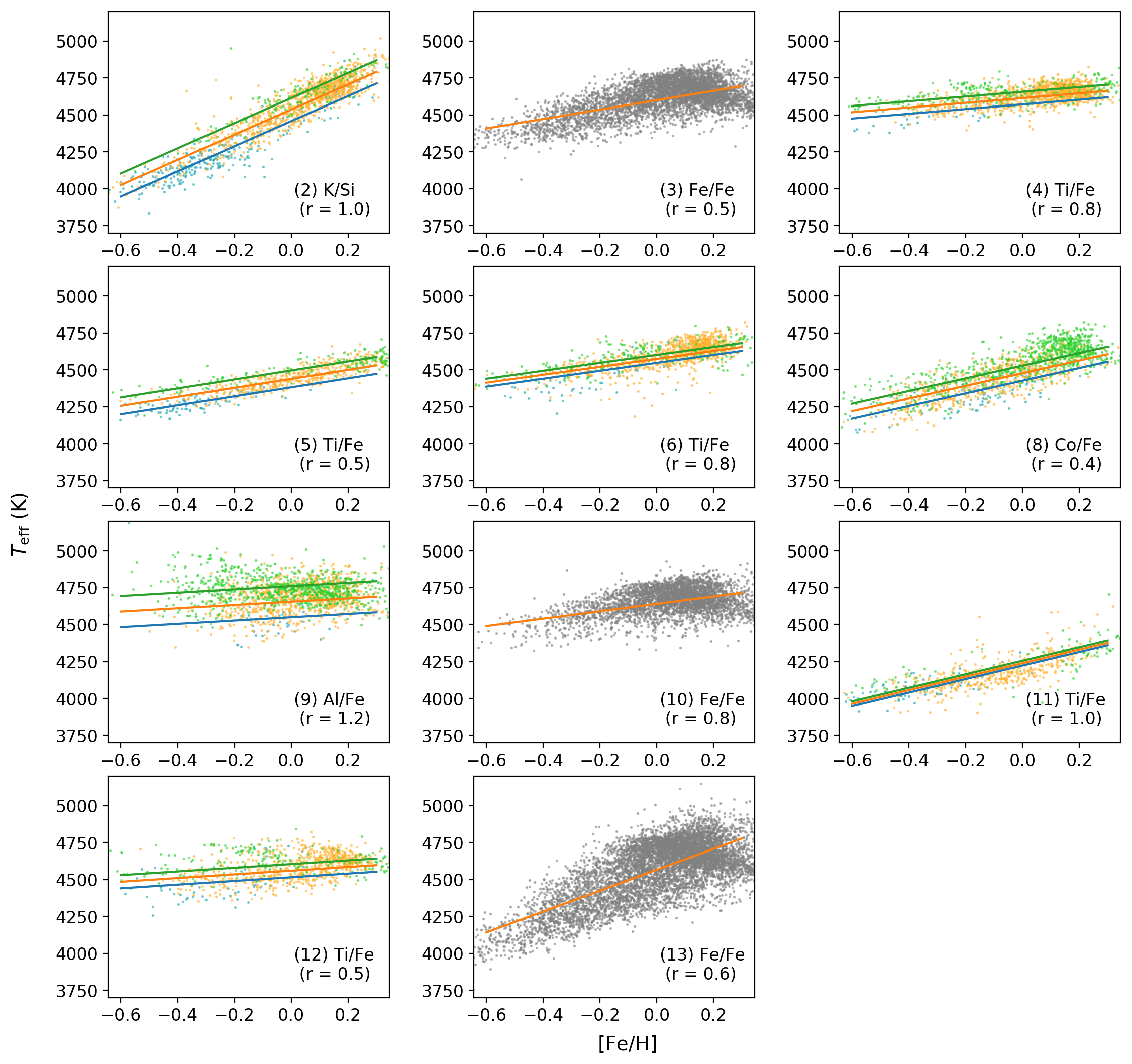}
		\caption{Each panel shows the variation of $\Teff$ at a fixed LDR ($r=r_0$) as a function of the metallicity $\FeH$ based on the relations in Table~\ref{tab:allfit}. The dots indicate the distribution of ($\Teff$, $\FeH$) for samples with $r$ within 0.05 of $r_0$. For line pairs in which $\XL$ and $\XH$ are different, the lines for different $\XLXH$ values are indicated in different colours: $-0.1$ (blue), $0$ (orange), and $0.1$ (green). Dots indicate the stars among our sample within $r = r_0 \pm 0.05$ and $\XLXH$ within $\pm 0.025$ of the value used for each relation; the colours of the dots indicate $\XLXH$, except for the pairs with two Fe lines.}
		\label{fig:zp}
	\end{figure*}

	For illustrating the size of the metallicity effect, Fig.~\ref{fig:zp} plots  $\Teff$ values at $r = r_0$ but with varying $\FeH$ and $\XLXH$ in equation~\ref{eq:wide}.
	$\Teff$ derived from a given LDR increases with increasing $\FeH$ and $\XLXH$. 
	All the line pairs show the dependency on $\FeH$, from ${\sim}100$\,K\,dex\textsuperscript{-1} to ${\sim}800$\,K\,dex\textsuperscript{-1}, which is in the same order of the model prediction discussed in \citet{fukue_line-depth_2015}.	
	The dependency on $\XLXH$ has various sizes depending on the elements involved; varying [Ti/Fe] gives a relatively small shift, from ${\sim}150$ to ${\sim}550$\,K\,dex\textsuperscript{-1}, while the dependency on [K/Si] and [Al/Fe] are larger, ${\sim}800$ and ${\sim}1000$\,K\,dex\textsuperscript{-1}, respectively.

\begin{figure}
		\includegraphics[width=\columnwidth]{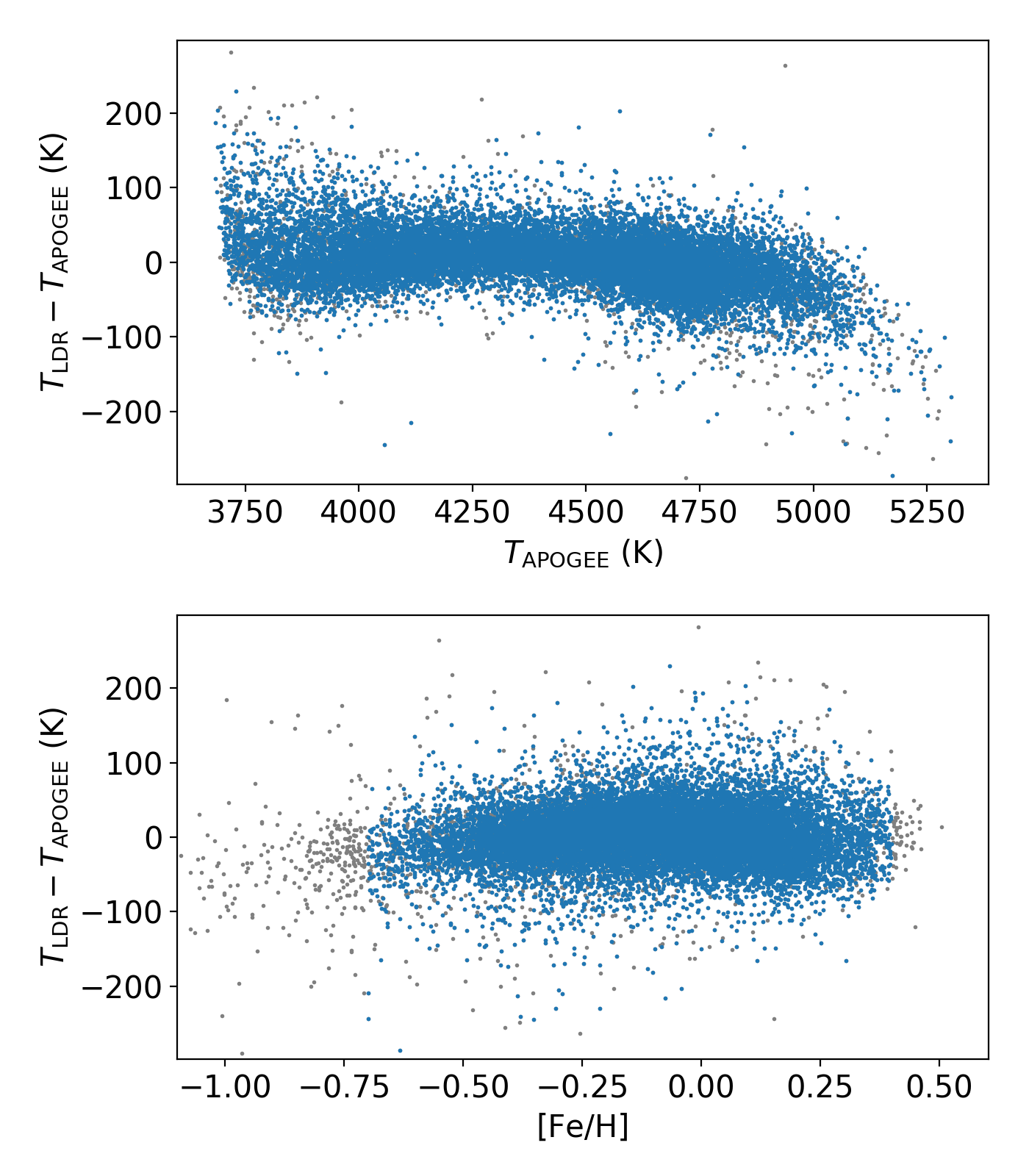}
		\caption{$T_\mathrm{LDR} - T_\mathrm{APOGEE}$ is plotted against $T_{\rm APOGEE}$ (upper panel) or $\FeH$ (lower panel). Blue points indicate stars with the metallicity and all abundances ratios within the boxes in Fig.~\ref{fig:XFe}, while grey points for others. The similar plots of the temperatures from the LDR relation of each line pair compared with $T_\mathrm{APOGEE}$ are given as Supporting Information.}
		\label{fig:ldrtta}
	\end{figure}

	We compare the temperatures derived by the LDR--$\Teff$--metal relations and those in the APOGEE catalogue, $T_\mathrm{APOGEE}$.
	For each star, we take a weighted average of temperatures from all available LDRs where the weights are given based on the residuals of the individual LDR--$\Teff$--metal relations (i.e. $\sigma$ in Table~\ref{tab:allfit}).
	This temperature based on the LDR method, $T_\mathrm{LDR}$, and $T_\mathrm{APOGEE}$ are consistent as presented in Fig.~\ref{fig:ldrtta}.
	At the both ends of the temperature range (lower than $3800$\,K and higher than $4800$\,K), the plot shows weak deviations from zero and slightly larger scatters. This may be due to imperfect formula we adopted, but adding the third-order term, $a'(r-r_0)^3$, didn't reduce the scatter significantly except for the line pair (5).
	Similarly to Fig.~\ref{fig:ldrtta}, the temperatures from the LDR relation of each line pair agree well with $T_\mathrm{APOGEE}$, and there are no particular ranges of temperature and metallicity where the differences between the LDR temperatures and $T_\mathrm{APOGEE}$ get commonly large (see Supporting Information).
	
	When we consider the entire sample within the black boxes in Fig.~\ref{fig:XFe}, the systematic offset between $T_\mathrm{LDR}$ and $T_\mathrm{APOGEE}$ is negligible, $2$\,K. 
	This is expected because $T_\mathrm{APOGEE}$ themselves were used for calibrating our relations. 
	The scatter of $35$\,K is smaller than the uncertainty given by \citealt{fukue_line-depth_2015} (${\sim}60$\,K).
	There is no metallicity term in the relations of \citet{fukue_line-depth_2015}.
    Among the 8 stars used for calibrating their LDR--$\Teff$ relations, 5 are outside the range of $-0.1 < \FeH < 0.1$, which can explain their larger scatter.  
	Fig.~\ref{fig:ldrtta} (b) shows that $T_\mathrm{LDR} - T_\mathrm{APOGEE}$ has no systematic trend with $\FeH$; the metallicity terms we introduced sufficiently remove the metallicity effect.
	
	As described in the webpage on APOGEE DR14 Stellar Abundances\footnote{http://www.sdss.org/dr14/irspec/abundances/}, there are warnings or caveats on abundances of some particular elements:
	$\mathrm{[Ti/H]}$ does not show the expected trend with metallicity (\citealt{holtzman_abundances_2015}; \citealt{hawkins_abundances_2016}).
	Co abundances in clusters show significant trends with $\Teff$, and a temperature-dependent calibration relation was applied to $\mathrm{[Co/H]}$, leaving a possibility of introducing a large systematic error.
	Although the coefficients of equation~\ref{eq:wide} may be affected by such systematics, the small standard deviations of the LDR--$\Teff$--metal relations for objects in the large parameter space manifest their advantage over the LDR--$\Teff$ relations.

	\section{Discussion}
	\label{sec:discussion}
	
    \subsection{Expected relation between the line depth and metallicity}
	Here we discuss a factor contributing to the metallicity effect, namely line saturation. 
	In general, the strength of an absorption line increases with metallicity. 
	The LDR would be independent of the metallicity if both low and high EP lines depend on the metallicity in the same manner. 
	Otherwise, the LDR would vary with the metallicity, which introduces the metallicity effect. 
	The relation between a line's equivalent width and the metallicity (to be exact, the number of absorbing atoms along the line of sight) is called the curve of growth, which is divided into three parts: linear, saturated and damped regions. 
    Here we consider the line depth instead of the equivalent width and discuss the curve of depth growth (hereafter CDG) which shows the growth of the line depth with increasing metallicity.
    The CDG is expected to show a similar trend with the curve of growth except the damped part for the following reason. 
	We confirmed by fitting the spectral lines with Gaussian profile that the widths of lines in the APOGEE spectra are independent of the line strengths and identical for all the relevant lines. 
    The full width at half maximum of the lines are on average $0.8$\,\AA~with the standard deviation of $0.1$\,\AA, and the line depth of $0.1$ typically corresponds to the equivalent width of $0.09$\,\AA.
    This linear relation between line depth and equivalent width is valid at least up to the depth of 0.6.
    Saturation occurs at around the depth of 0.1, but the line depth can trace the line growth even in the saturated region because of the line broadening including macroturbulent and instrumental ones (the instrumental broadening is dominant in APOGEE spectra of giants).
    Absorption at neighbouring wavelengths around the line centre keeps growing in the saturated region and contributes to the observed depth at the centre thanks to the broadening; the CDG therefore shows the growth in the saturated region like the curve of growth.
    In contrast, growth in the damped part is expected to give little effect on the depth of the CDG when spectral resolution is high enough to resolve the damping wing. 
    Note that we changed atmosphere models according to the metallicity in our CDG plots. 
    This may cause an additional change in the line growth, but our simulation indicates that the slope of the CDG is still unity in the linear part.

	\subsection{Match between the CDG and the observational data}
	Whether the LDR is affected by the metallicity or not depends on which regions of the CDG the lines lie in.
	When both lines are in the linear region, they get deeper in the same way with increasing the metallicity and thus the metallicity effects on the two depths cancel each other out when the ratio is taken. 
	This is actually the reason why weak lines are suggested to use by some authors (e.g. \citealt{gray_spectral_1994}, \citealt{kovtyukh_determinations_2006}). 
	In contrast, when lines are saturated, the line depths may show different responses to the metallicity depending on the degree of saturation. 
	In such a case, the LDR depends on the metallicity.
	
	To check whether the line saturation actually affects our LDRs, we compare the observed line depths with theoretical CDGs.  
    In the following discussion, $\FeH$ indicates the metallicity.
    All the abundance ratios including $\mathrm{[\alpha/Fe]}$ are fixed to 0, and $\mathrm{[X/H]}$ of any heavy element changes following the change of $\FeH$. 
    
	We used synthetic spectra produced by MOOG \citep{MOOG} making use of ATLAS9 stellar atmosphere models from \citet{meszaros_new_2012} and the line list of APOGEE \citep{shetrone_sdss-iii_2015} to calculate theoretical LDRs. 
    Both atomic lines and molecular lines\footnote{The line list of \citet{shetrone_sdss-iii_2015} includes lines of six molecules, $\mathrm{H_2}$, $\mathrm{OH}$, $\mathrm{C_2}$, $\mathrm{CN}$, $\mathrm{CO}$, and $\mathrm{SiH}$. However, no $\mathrm{SiH}$ line exists around the lines we considered for LDRs.} are included in the synthetic spectra.
    With the help of the synthetic spectra, we can investigate the metallicity effect in a wider parameter space than one can explore only with observational data.
	Including low-metal regions allows us to examine the linear part of the CDG and to determine whether the observational data are affected by the line saturation.
    We changed $\logg$ and microturbulence of the models considering the relations between these two parameters and $\Teff$ found in the APOGEE catalogue as illustrated in Fig. \ref{fig:FeH}.
	The orange points in Fig.~\ref{fig:FeH} indicate the stellar parameters we used to calculate synthetic spectra.
	Because we use the atmospheric models from \citet{meszaros_new_2012} with the grid spacing of 0.5 dex in $\logg$, the adopted $\logg$ are slightly biased from the sequence of the APOGEE sample, especially at $\Teff \ge 4750$\,K.  
	However, at this temperature range, changing $\logg$ by 0.5 dex would change the predicted LDR by less than 0.03 in the entire metallicity range and less than 0.01 in the range of the observed sample we consider, $-0.7 < \FeH < 0.4$\,dex.
	The small offsets in $\logg$ have therefore no significant impact on the conclusions in this paper.

	\begin{figure}
		\centering
		\includegraphics[width=\columnwidth]{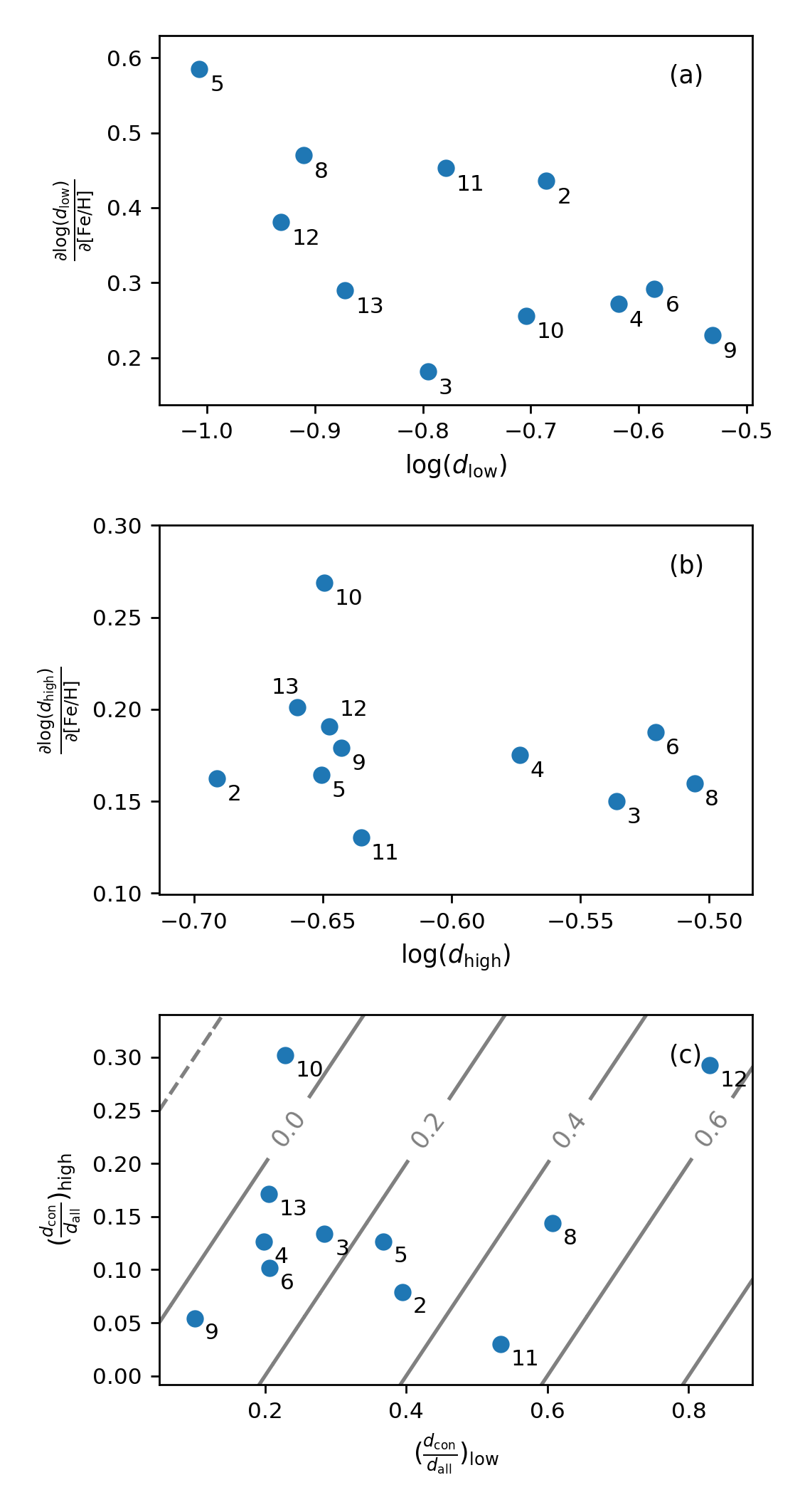}
		\caption{In the panels (a) and (b), the slopes ($\frac{\partial\log{d}}{\partial \FeH}$) of the CDGs are plotted against the depths ($\log{d}$) of low EP (a) and high EP (b) lines at $\Teff=4500$\,K and $\FeH=0.0$\,dex. The panel (c) plots the contamination, $\frac{d_\mathrm{con}}{d_\mathrm{all}}$, where $d_\mathrm{con}$ indicates the depths in the synthetic spectra with all lines except the target line included and $d_\mathrm{all}$ indicates those with all lines included. The fractional change in LDR following the $d_\mathrm{con}/d_\mathrm{all}$ values is illustrated by grey contours. The IDs of the line pairs are indicated within each panel.}
		\label{fig:slope}
	\end{figure}    

    The depth and slope of each line in the CDG based on synthetic spectra at $\Teff = 4500$\,K and $\FeH = 0.0$\,dex are plotted in Fig.~\ref{fig:slope}.
    The line is in the linear region if the slope of the CDG ($\partial \log{d}/\partial\FeH$) is 1. 
    It is clear that all the lines are saturated in this temperature and metallicity.
    We note that the line pairs (10) and (12) have contaminations larger than other line pairs as seen in Fig.~\ref{fig:slope}~(c).
    While the characteristics of the LDR relations of these two pairs such as the scatters around the relations seem similar to other line pairs, the strong blends may introduce large errors under some conditions.
    Fig.~\ref{fig:slope} also suggests that the characteristics of the line pair (4) is not particularly different from those of the others.
    Thus the line pair (4), Ti I $15543.78$\,\AA/Fe I $15591.49$\,\AA, is considered as an example in the following discussion unless otherwise mentioned.
	
	\begin{figure}
		\centering
		\includegraphics[width=0.6\columnwidth]{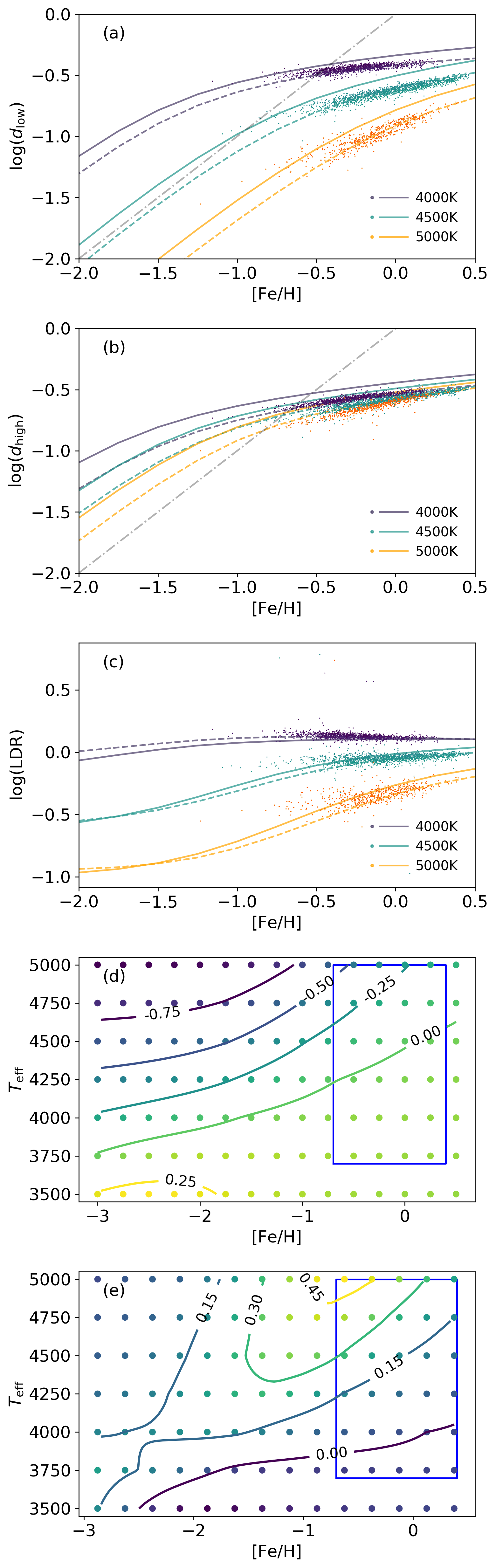}
		\caption{Theoretical prediction of some characteristics of the line pair (4), Ti I $15543.78$\,\AA~and Fe I $15591.49$\,\AA, based on synthetic spectra. The top panels plot the CDGs of the low--EP Ti I line (panel a) and those of the high--EP Fe I line (panel b) at three temperatures, while the panel (c) plots the LDRs at the same temperatures as functions of the metallicity. In these three panels, the predictions made by synthetic spectra with only LDR lines included ($d_\mathrm{t}$ and $\mathrm{LDR_t}$, indicated by dashed curves) are compared with the counterparts with all absorption lines included in the synthesis ($d_\mathrm{all}$ and $\mathrm{LDR_{all}}$, indicated by solid curves). In addition, the distributions of the objects among our sample with $\Teff$ within $\pm 50$\,K of the three temperatures are indicated by dots. The panels (d) and (e) respectively shows how the LDR and the metallicity effect, $\frac{\partial \log{r}}{\partial \FeH}$, depend on $\FeH$ and $\Teff$. The synthetic spectra with all lines included were calculated for a grid of ($\FeH$, $\Teff$), and the LDR and metallicity effect based on them are illustrated by the colour of dots on the grid points and the contours. The blue boxes inside these panels indicate the temperature and metallicity ranges of APOGEE red giants we used for the LDR relations.}
		\label{fig:LP4}
	\end{figure}
	
	Fig.~\ref{fig:LP4} shows that the theoretical CDGs match the observational trend but show systemic offsets. 
    The difference between the theoretical CDGs and the measured LDRs may be due to the determination of continuum level, inaccurate $\log{gf}$ values, and/or variations of $\TiFe$ and other ratios from the solar values that we assumed for the synthetic spectra. 
    The continuum in the final spectra may depend on the algorithm and the tool used for the normalisation.
    We made a comparison between the line depths of the line pair (4) in synthetic spectra normalised by the IRAF \textsc{continuum} task and those in synthetic spectra themselves (the theoretically true continuum).
    The depth differences between the former spectra and those in the latters are smaller than 10\%.
    This discrepancy has a weak effect on the LDR, $|\Delta \log{r}| < 0.03$, regardless of the line depth.
    Therefore, the continuum normalisation have a very small effect on the LDR-$\Teff$ relations.
    However, there may well be other factors which affect the normalisation of real observed spectra, e.g., cosmic artifacts, data-reduction residuals, and spectral resolution variation. 
    The uncertainties in continuum normalisation may contribute to the scatter of our derived LDR--$\Teff$--$\FeH$ relations.
    It is advised to use the same normalisation method in both the calibration phase of constructing the relations and in the application phase of estimating $\Teff$ of new targets especially when spectral resolution is not high.
	Nevertheless, the theoretical CDGs reproduce the shapes of the observational CDGs well, which enables us to conclude that the line saturation is the main course of the metallicity effect as follows.  

	In Fig.~\ref{fig:LP4}, the CDGs for the low EP line indicate that the linear region appears at $\FeH < -1.5$\,dex with $\Teff \sim 4000$\,K, while they are almost totally in the linear region with $\Teff > 5000$\,K. 
	For the high EP line, the CDGs are found to be in the saturated region at the metallicity around $-1.5$\,dex and higher. 
	Synthetic spectra of other line pairs also show similar trends; this result indicates that all high EP lines measured in our sample are saturated as well as low EP lines at $\Teff \le 4500$\,K. 
	In contrast, some low EP lines remain in the linear region at relatively high temperatures. 
	With the different slopes of CDGs, the LDR changes with $\FeH$ even at a fixed $\Teff$ (Fig.~\ref{fig:LP4}, panel d). 
    Therefore the line saturation, at least partly, explains the metallicity effect in the LDR--$\Teff$--metal relations.
	As illustrated in Fig.~\ref{fig:LP4} (e), the metallicity effect gets small in two different cases: very low metallicity ($\FeH \lesssim -1.5$\,dex) and a narrow strip at a low temperature ($\Teff \,{\sim} 4000$\,K; see the almost flat sequence of the purple dots in Fig.~\ref{fig:LP4} (c)).
	In the former case, both lines are in the linear region, while in the latter case the CDGs of both low and high EP lines have similar slopes in the saturated region.
	In fact, in equation~\ref{eq:wide}, the dependency on the metallicity, $d\FeH + e\FeH(r-r_0)$, becomes close to zero with $r \,{\sim} 0.8$ ($\Teff \,{\sim} 4000$\,K) for the line pair (4).
    
    \subsection{The effect of line blends}
    \label{sec:line_blend}
    It is worthwhile to discuss the effects of blends.
    We used MOOG synthetic spectra to measure the degree of contamination by molecular and atomic lines to each target line. 
    Three kinds of synthetic spectra were produced by using different sets of lines: (1) the target line only, (2) all lines except the target line, and (3) all lines. 
    Then, the line depths (or LDRs) are measured in the three spectra and labelled as $d_\mathrm{t}$, $d_\mathrm{con}$ and $d_\mathrm{all}$ (or $\mathrm{LDR_t}$, $\mathrm{LDR_{con}}$ and $\mathrm{LDR_{all}}$), respectively.
	As Fig.~\ref{fig:slope} (c) indicates, the contamination fraction, $d_\mathrm{con} / d_\mathrm{all}$, is significant, 10\%--60\%, for many line pairs. 
	Such contaminations are similar to what was found by \citet{fukue_line-depth_2015}. 
	However, our simulation also indicates that the contaminations of low and high EP lines in each pair for various $\Teff$ and $\FeH$ tend to be similar. 
    For example, $d_\mathrm{t}$ is found to be slightly smaller than $d_\mathrm{all}$ for the line pair (4) as presented in the panels (a) and (b) in Fig.~\ref{fig:LP4}, but their LDR values are similar, as shown in the panel~(c) in Fig.~\ref{fig:LP4}. 
	In fact, the impact of contamination on our LDR values is relatively small, less than 20\% for about half of the line pairs as shown in Fig.~\ref{fig:slope}~(c). 
    Besides, the metallicity effect on the LDR does exist even in a simulation using artificial spectra with only target lines included (i.e., no blends at all).
    This further supports that the line saturation is the main factor which causes the metallicity effects we detected even if blends or other factors could contribute to the metallicity effects for some line pairs.
    If the contaminations depend only on $\Teff$ and metallicity, well calibrated LDR--$\Teff$--metal relations can predict $\Teff$ precisely. 
    Nevertheless, some parameters which are not taken into account in our analysis may affect the contaminations and then contribute to scatters around the relations.  
    
	\section{Summary}
	\label{sec:summary}
	
	By using the large dataset of the APOGEE DR14, the LDR method with $H$-band spectra was revisited to investigate the metallicity effect.
	Seven of the line pairs in \citet{fukue_line-depth_2015} were re-calibrated by about more than 1,000 spectra of solar-metal stars and found to be consistent with their relations within the larger errors in the previous study.
	Four new line pairs which show similarly good correlations between $\Teff$ and LDR were also found.
	The 11 line pairs were then investigated with ${\sim}17,000$ spectra in a wide range of metallicity to obtain LDR--$\Teff$--metal relations, all of which have significant metallicity-terms ($ {\sim}100$ to ${\sim}800\,\mathrm{K\,dex^{-1}}$) and abundance-dependent terms ($ {\sim}150$ to ${\sim}1000\,\mathrm{K\,dex^{-1}}$).
	Making use of synthetic spectra for examining how the metallicity affects two absorption lines in a given LDR pair, we found that most of the lines measured in our sample are saturated, which can explain the metallicity effect on the LDRs.
	With the metallicity effect taken into account, our LDR--$\Teff$--metal relations give $\Teff$ consistent with APOGEE within the standard deviation of $35$\,K.
	
	\section*{Acknowledgement}
	The authors acknowledge very useful comments from the referee.
    We are also grateful to Daisuke Taniguchi and Naoto Kobayashi for discussions.
	This work was supported by a Grant-in-Aid, KAKENHI, from the Japan Society for the Promotion of Science (JSPS; No.~26287028).
    N.M. is supported by a JSPS KAKENHI (No. 18H01248).
	K.F. also acknowledges the JSPS for KAKENHI for Research Activity Start-up (No.~16H07323).
	
	
	
	\bibliographystyle{mnras}
	\bibliography{LDR-cite.bib} 
	
	\section*{Supporting Information}
    Additional Supporting Information may be found in the online version of this article:

    \textbf{Figure4\_additional\_plots.pdf}, LDR--$\Teff$ relations for samples within different abundance ranges of all the line pairs, similarly to Fig.~\ref{fig:r1}. 
    
    \textbf{Figure6\_additional\_plots.pdf}, the temperatures from the LDR relation of each line pair are compared with $T_\mathrm{APOGEE}$, similarly to Fig.~\ref{fig:ldrtta} where $T_\mathrm{LDR}$ obtained with all available LDR relations are used.
	
	\bsp	
	\label{lastpage}
\end{document}